\documentclass[12pt]{article}

\begin{document}

\title{Gage-equivalent forms of the Schr\"odinger equation for a hydrogenlike atom in a time-dependent electric field}

\author{Yuri~V. Popov\\ {\small \it Institute of Nuclear Physics, Moscow State University, Moscow 119992, Russia}\\
{\small e-mail: popov@srd.sinp.msu.ru}\\ \\Konstantin~A. Kouzakov\\
{\small \it Faculty of Physics, Moscow State University, Moscow 119992, Russia}\\
{\small e-mail: kouzakov@srd.sinp.msu.ru}}

\date{}
\maketitle


\vskip 12pt


\begin{abstract}\noindent Several gage-equivalent forms (including some novel ones) of the Schr\"o-dinger
equation for a hydrogenlike atom in a time-dependent electric
field of a laser pulse are presented. These forms allow to develop
a perturbation theory for both small and rather large intensities
of the electromagnetic field.
\end{abstract}

\section{Introduction}
Exploring the interaction between the electromagnetic field and matter is the major and oldest issue of both classical
and quantum physics. It has a number of branches. One of them is the interaction of a strong laser field with atoms and
molecules. Here the concept of a ``strong field'' implies the field which is comparable with an electric field in an
atom, which binds electrons to nucleus. The strong field may result in single or multiple ionization of an atom or
molecule. By studying the energy and angular distributions of the ionized electrons physicists expect to obtain
information about the structure of a quantum object and the ionization mechanisms.

The coming into being of the theory of such processes is due to pioneering works of Keldysh~\cite{Keldish} and his
followers. Since then, a number of studies have appeared analyzing pluses and minuses of this theory, investigating its
limits and suggesting corrections for its improvement. Just citing these works would take a half of the journal volume
and therefore we cite only several recent review papers~\cite{Obzori1,Obzori2,Obzori3,Obzori4,Obzori5,Obzori6}, where
trends in this field of science can be seen. However, the theoretical and mathematical content of the majority of works
in recent years has shifted towards development of numerical schemes for solving basic equations. This is due to the
intense growth of computational facilities of modern computers.
Despite the enormous progress of such approach in understanding
the processes taking place under the action of an intense laser
pulse, the analytical models remain of very current importance,
since they have the power of prediction. The exact solution of the
considered problem is known only for a very small set of local
potentials, in particular, for an oscillating potential (see, for
instance,~\cite{oscillator1,oscillator2}). However, there is no
ionized states in this potential. For the simplest practical case,
namely for the hydrogen atom, one already must consider different
approximations whose mathematical correctness is not always clear.

In this connection the property of gage invariance of the
electromagnetic field is often helpful. In turn, this property
allows to obtain various equivalent forms of the time-dependent
Schr\"odinger equation (TDSE) related to each other through
unitary transformations which, as is known, lead to the invariance
of the physical quantities given by quadratic forms of the wave
function. Let us recall that the Maxwell equations can be written
in terms of the scalar and vector potentials, $U(\vec r,t)$ and
$\vec A(\vec r,t)$. These potentials quite unambiguously determine
the observed characteristics of the electromagnetic field, namely
the electric and magnetic field intensities, $\vec E$ and $\vec
H$. At that, the potentials themselves are defined ambiguously.
For example, two sets of potentials $(\vec A', U')$ and $(\vec A,
U)$, where
$$
\vec A'=\vec A + \vec\nabla f, \qquad U'=U-\frac1c\frac{\partial
f}{\partial t},
$$
give the same electric and magnetic field intensities for an
arbitrary function $f(\vec r,t)$.

The forms of TDSE using different gage transformations of the electromagnetic field and some of the corresponding
useful consequences are the subject of this work. The atomic units $e=m_e=\hbar=1$ are used throughout. According to
this system of units the light velocity $c$ is about 137.

\section{Coordinate representation}
First of all it should be noted that the field intensity is, upon definition, related to the
potential as follows:
$$
\vec E(\vec r,t)=-\frac 1c \frac{\partial}{\partial t} \vec A(\vec
r,t)- \vec\nabla U(\vec r,t). \eqno (1)
$$
In the so-called Coulomb gage it is assumed that
$$
{\rm div}\vec A=0.
$$
In the simplest study one makes a physical assumption about a weak
dependence of the scalar potential on the coordinate within the
atom, i.e.
$$
U(\vec r,t)\simeq U(0,t),
$$
which allows to neglect the gradient of the scalar potential in (1). This leads to a well known
dipole approximation
$$
\vec A(\vec r,t) \simeq \vec A(0,t) = \vec A(t).
$$
Setting $\vec A(t)=\vec e A(t)$, where $\vec{e}$ is the unit
polarization vector, we obtain a linearly polarized laser beam.
The condition of the absence of the field outside the time
interval $(0,T)$, where the laser pulse acts, takes the form
$A(t\leq 0)=A(t\geq T)=0$.

Consider the TDSE, which describes the interaction between an
electric pulse and a hydrogenlike atom,
$$
\left\{i\frac{\partial}{\partial t} -\frac12\left[-i\vec\nabla+\frac{1}{c}\ \vec
eA(t)\right]^2+\frac{Z}{r}\right\}\Psi(\vec r,t)=0, \qquad \Psi(\vec r,0)=\sqrt{\frac{Z^3}{\pi}}e^{-Zr}, \eqno (2)
$$
where $Z$ designates the nuclear charge. The initial state of the problem allows to conclude that
at any time moment $t$ we deal with a square integrable wave packet. Moreover, a normalization
condition should be fulfilled:
$$
\int d\vec r |\Psi(\vec r,t)|^2=1, \eqno (3)
$$
whose physical meaning is a conservation of the total probability of all events in the system.

The well known unitary transformation
$$
\Psi(\vec r,t)=\exp{\left[-i\frac1c A(t) (\vec e\vec
r)\right]}\Phi_L(\vec r,t)
$$
results in the following form of TDSE:
$$
\left[i\frac{\partial}{\partial t}+ \frac12\triangle - \vec
E(t){\vec r}+\frac{Z}{r}\right]\Phi_L(\vec r,t)=0, \qquad
\Phi_L(\vec r,0)=\Psi(\vec r,0). \eqno (4)
$$
Here $\vec E(t)=-\vec e \partial A(t)/c\partial t$. The notation
$\Phi_L(\vec r,t)$ indicates the so-called length form of TDSE. In
this context, $\Psi(\vec r,t)\equiv\Phi_V(\vec r,t)$ is sometimes
referred to as the velocity form. Usually one requires a good
numerical algorithm to give a coincidence (within the accuracy) of
the computed observed quantities (the level occupations, angular
and energy distributions of the ionized electrons and etc.) in the
length and velocity forms. In the exact theory they must be
identical.

The less known Henneberger-Kramers transformation employs the unitary operator of the space
shift~\cite{Kramers}:
$$
\Psi(\vec r,t)=\exp\left[b(t)(\vec e\vec\nabla)-\frac{i}{2c^2}\int\limits_0^tA^2(\tau)d\tau\right]\Phi_{HK}(\vec r,t),
\quad b(t)=-\frac{1}{c}\int\limits_0^tA(\tau)d\tau. \eqno (5)
$$
Inserting (5) into (2) and setting for convenience that $b(t)=Af(t)$, where $|f(t)|\leq 1$ and
$f(t\leq 0)= f(t\geq T)=0$, we get
$$
\left[i\frac{\partial}{\partial t}+\frac12\triangle +\frac{Z}{|\vec r-\vec e Af(t)|}\right]\Phi_{HK}(\vec r,t)=0,
\qquad \Phi_{HK}(\vec r,0)=\Psi(\vec r,0).
 \eqno (6)
$$
Note that at any time moment $t$ the wave packet is normalized to unity:
$$
\int d\vec r |\Phi_{HK}(\vec r,t)|^2=1.
$$

Making the following scaling transformation:
$$
t=A\tau, \qquad \vec r=A\vec x, \qquad \Phi_{HK}(\vec r,t)=
A^{-3/2}\phi(\vec x,\tau),
$$
we obtain
$$
\left[iA\frac{\partial}{\partial\tau}+\frac12\triangle_{x}
+\frac{AZ}{|\vec x-\vec e f(A\tau)|}\right]\phi(\vec x,\tau)=0.
\eqno (7)
$$
Let us set
$$
\phi(\vec x,\tau)=Ne^{-AS(\vec x, \ \tau)}, \eqno (8)
$$
where $\Re(S)>0$ if $x\to\infty$. It follows from (8) that
$$
\left[i\frac{\partial S}{\partial\tau}-\frac12(\vec\nabla_x S)^2
\right]+\frac{1}{A}\left[\frac12\triangle_x S -\frac{Z}{|\vec
x-\vec e f(A\tau)|}\right]=0. \eqno (9)
$$

In the absence of an external electric field, $f(t)=0$ and
$$
S_0(\vec x, \tau)=Zx-\frac{i}{2}Z^2\tau.
$$
This function satisfies not only (9) but also the equation
$$
\left[i\frac{\partial S_0}{\partial\tau}-\frac12(\vec\nabla_x
S_0)^2 \right]=0.
$$
This fact allows to use the perturbation series if $A\gg1$:
$$
S(\vec x, \tau)=\sum\limits_{n=0}\left(\frac{1}{A}\right)^n
S_n(\vec x, \tau). \eqno (10)
$$
For instance, the term $S_1$ satisfies the linear nonhomogeneous partial differential equation
$$
i\frac{\partial S_1}{\partial\tau}-\left(\frac{\vec x}{x} \vec\nabla_x\right) S_1 +
\left[\frac{Z}{x} -\frac{Z}{|\vec x-\vec e f(A\tau)|}\right]=0, \eqno (11)
$$
whose particular solution is
$$
S_1(\vec x,\tau)=i\int\limits_0^{\tau} d\xi\left[\frac{Z}{i\xi +
C}-\frac{Z}{|(i\xi + C)\vec x/x-\vec ef(A\xi)|}\right]. \eqno (12)
$$
In formula (12), $C=x-i\tau$ is an integral of motion of equation (11), in which connection
$$
|(i\xi + C)\vec x/x-\vec ef(A\xi)|=\sqrt{(i\xi + C)^2 + f^2(A\xi)-2\frac{(\vec x\vec e)}{x}(i\xi +
C)f(A\xi)}.
$$
After the termination of the laser pulse ($t\geq T)$
$$
S_1(\vec x,\tau)=i\int\limits_0^{T/A} d\xi\left[\frac{Z}{i\xi +
C}-\frac{Z}{|(i\xi + C)\vec x/x-\vec ef(A\xi)|}\right]. \eqno (13)
$$

For the term $S_2$ we have, in accordance with (9) and (10), the equation
$$
\left[i\frac{\partial S_2}{\partial\tau}-\left(\frac{\vec
x}{x}\vec\nabla_x\right) S_2 \right] + \frac12\left[\triangle_x
S_1 -(\vec\nabla_x S_1)^2\right]=0. \eqno (14)
$$
Correspondingly, its solution is
$$
S_2(\vec x,\tau)=\frac{i}{2}\int\limits_0^{\tau} d\eta
\left[\triangle_x S_1 -(\vec\nabla_x S_1)^2\right]. \eqno (15)
$$
To apply the gradient and Laplas operators to the function
$S_1(\vec x,\eta)$ in (15), at first it is necessary to set
$C=x-i\tau$ in (12), then to perform these differential
operations, and after that to make the substitution $\vec
x=(i\eta+C)\vec x/x$.

In (13) and (15) one can return back to the variables $(\vec r,t)$
and see that the argument of the exponent in (8) does not
explicitly depend on $A$. This is a footprint of a quasi-classical
approximation.

For estimating the value of $A$ we consider a particular case of
the laser pulse shape which is frequently utilized in
calculations:
$$
A(t)=\left\{ \begin{array}{ll} \displaystyle
A_0\sin^2(\pi{t}/{T})\sin(\omega t+\varphi) \quad & (0\leq t\leq
T),
\\[3mm] \displaystyle A(t)=0 \qquad & (t\geq T), \end{array} \right.
\quad \frac{A_0}{c}=\frac{1}{\omega}\sqrt{\frac{I}{I_0}}. \eqno
(16)
$$
In (16), $I_0=3.5\times 10^{16}$ Wt/cm$^2$ is the unit of the field intensity in an atom,
$\omega=0.056$ (the base frequency of the titan-sapphire laser), $T\approx 2\pi n/\omega$, and $n$
is a number of cycles in the pulse. The phase $\varphi$ must be chosen according to the condition
$b(T)=0$. Setting $n=10$ and $I\sim 10^{14}$ Wt/cm$^2$, we obtain the estimate $A\sim 20$, or
$1/A\sim 0.05$. This allows to expect a good convergence of the series (10) in the case of a rather
strong field with moderate carrier frequency. In this range the experimental data have been
obtained which allow to check the correctness of the derived expansion in the reversed powers of
the field intensity.

\section{Momentum representation}
TDSE for the considered problem in momentum space follows from (2)
and has the form
$$
\left\{i\frac{\partial}{\partial t}-\frac12\left[\vec p+\frac{1}{c}A(t)\vec e \right]^2 \right\}\tilde{\Psi}(\vec
p,t)+\int\frac{d^3 p'}{(2\pi)^3}\frac{4\pi Z}{|\vec p-{\vec p}'|^2}\tilde{\Psi}({\vec p}',t)=0, \eqno (17)
$$
$$
\tilde{\Psi}(\vec p,0)=\frac{8\sqrt{\pi Z^5}}{(p^2+Z^2)^2}.
$$
In equation (17) the function $\tilde{\Psi}(\vec p,t)$ designates the Fourier transform of the
function ${\Psi}(\vec r,t)$. The unitary transformation
$$
\tilde{\varphi}(\vec p,t)=e^{i\frac{t}{2}p^2-ib(t)(\vec e\vec
p)+i\frac12\int\limits_0^t d\tau [b'(\tau)]^2}\tilde{\Psi}(\vec
p,t) \eqno (18)
$$
leads to the equation
$$
i\frac{\partial}{\partial t}\tilde{\varphi}(\vec
p,t)+\frac{Z}{2\pi^2}\int\frac{d^3x}{x^2}
e^{-i\frac{t}{2}x^2+i[t\vec p-b(t)\vec e]\vec x}
\tilde{\varphi}(\vec p-\vec x,t)=0. \eqno (19)
$$
One can obtain the analogous equation in coordinate space upon making the Fourier transform
$$
\tilde{\varphi}(\vec p,t)=\int d^3r \ e^{-i\vec p\vec
r}{\varphi}(\vec r,t).
$$
In this case we obtain from (19)
$$
i\frac{\partial}{\partial t}{\varphi}(\vec
r,t)+\frac{Z}{2\pi^2}\int\frac{d^3x}{x^2}
e^{i\frac{t}{2}x^2+i[\vec r-b(t)\vec e]\vec x} {\varphi}(\vec
r+t\vec x,t)=0. \eqno (20)
$$

Using the momentum shift operator, equation (19) can be presented
in the form
$$
i\frac{\partial}{\partial t}\tilde{\varphi}(\vec
p,t)+\frac{Z}{2\pi^2}\int\frac{d^3x}{x^2}
e^{-i\frac{t}{2}x^2+i[t\vec p-b(t)\vec e]\vec x} e^{-\vec
x\vec\nabla_p}\tilde{\varphi}(\vec p,t)=0. \eqno (21)
$$
The Weyl operator identity leads to the following result:
$$
e^{i\vec p\vec x t } e^{-\vec x\vec\nabla_p}\equiv
e^{i\frac{t}{2}x^2} e^{i\vec x(\vec p t+i\vec\nabla_p)},
$$
which allows to obtain from (21) the equation
$$
i\frac{\partial}{\partial t}\tilde{\varphi}(\vec
p,t)+\frac{Z}{2\pi^2}\int\frac{d^3x}{x^2} e^{i\vec x\vec
\mathcal{H}} \tilde{\varphi}(\vec p,t)=0. \eqno (22)
$$
Here $\vec\mathcal{H}=t\vec p-b(t)\vec e+i\vec\nabla_p$. Equation
(22) can be presented in a more compact operator form, upon
integrating over $\vec x$:
$$
i\frac{\partial}{\partial t}\tilde{\varphi}(\vec
p,t)+\frac{Z}{|t\vec p-b(t)\vec e
+i\vec\nabla_p|}\tilde{\varphi}(\vec p,t)=0, \qquad
\tilde{\varphi}(\vec p,0)=\tilde{\Psi}(\vec p,0). \eqno (23)
$$
A similar equation can be obtained for ${\varphi}(\vec r,t)$ as well:
$$
i\frac{\partial}{\partial t}{\varphi}(\vec r,t)+\frac{Z}{|\vec
r-b(t)\vec e-it\vec\nabla_r|} {\varphi}(\vec r,t)=0, \qquad
{\varphi}(\vec r,0)={\Psi}(\vec r,0). \eqno (24)
$$
The eigenfunctions of the operator $\vec\mathcal{H}$ are the Volkov states~\cite{volkov}
$$
\chi(\vec p,t)=e^{i\frac{t}{2}p^2-ib(t)(\vec e\vec p)-i\vec p\vec r },
$$
i.e. $f(\vec\mathcal{H})\chi=f(\vec r)\chi$. Expanding the function $\tilde{\varphi}(\vec p,t)$
over the basis of the Volkov states, we again obtain (17).

Thus, we arrive at the operator equation
$$
\frac{\partial\mathcal{S}(t)}{\partial
t}=iZ\mathcal{A}(t)\mathcal{S}(t), \qquad \mathcal{S}(0)=I, \qquad
\tilde{\varphi}(\vec p,t)=\mathcal{S}(t)\tilde{\varphi}(\vec p,0),
\eqno (25)
$$
with $\mathcal{A}(\tau)=1/|\vec\mathcal{H}|$. Its formal solution
can be presented, for example, in the form of the Magnus
expansion~\cite{magnus}
$$
\mathcal{S}(t)=\exp\left[iZ\int\limits_0^t
d\tau\mathcal{A}(\tau)\right]=\exp\left(\sum_{n=1}^\infty
B_n\right), \eqno (26)
$$
where
$$
B_1=iZ\int\limits_0^t d\tau\mathcal{A}(\tau),
$$
$$
B_2=-\frac{Z^2}{2!}\int\limits_0^t d\tau_1\int\limits_0^{\tau_1}
d\tau_2[\mathcal{A}(\tau_1),\mathcal{A}(\tau_2)],
$$
$$
B_3=-\frac{iZ^3}{3!}\int\limits_0^t d\tau_1\int\limits_0^{\tau_1} d\tau_2\int\limits_0^{\tau_2}
d\tau_3\left\{[\mathcal{A}(\tau_1),[\mathcal{A}(\tau_2),\mathcal{A}(\tau_3)]]+
[[\mathcal{A}(\tau_1),\mathcal{A}(\tau_2)],\mathcal{A}(\tau_3)]\right\}
$$
and so on. This leads to necessity of calculating commutators of
the operator $\mathcal{A}(t)$ at different time moments $t$.

Despite the aesthetic attraction and symmetry of equations in the
$(\vec r,\vec p)$ variables, one can use it only in the context of
the perturbation theory with respect to the reverse powers of the
parameter $A$ (see (6)). One can obtain the exact solutions of
equation (25), if the operator $\mathcal{A}(t)$ is expressed as a
finite linear combination of the Lie algebra generators with
time-dependent coefficients. As to the solution in the form of the
Magnus expansion (26), its utilization seems to be not very
efficient in practice due to unclear physical interpretation of
the operators $\exp(B_i)$, as opposed, for example, to the cases
of the space and momentum shift operators.

\vskip 12pt

We would like to express our gratitude to Profs. A.~V.~Mikhalev and V.~F.~Butuzov for useful
discussions and remarks.


\begin{thebibliography}{99}

\bibitem{Keldish} Keldysh L. V. Ionization in the Field of a Strong Electromagnetic Wave
// Sov. Phys. JETP. -- 1965. -- Vol. 47, no. 5. -- P. 1307--1314.
%
\bibitem{Obzori1} Delone N.~B., Krainov V.~P. Tunnelling and barrier-suppression ionization of atoms and ions in a laser radiation field
// Physics-Uspekhi. -- 1998. -- Vol. 41, no. 5. -- P. 469--485.
%
\bibitem{Obzori2} Lambropoulos P., Maragakis P., Zhang J. Two-Electron Atoms in Strong Fields // Phys. Rep. -- 1998.
-- Vol. 305, no. 5. -- P. 203--293.
%
\bibitem{Obzori3} Gavrila M. Atomic stabilization in superintense laser fields
// J. Phys. B.: At. Mol. Opt. Phys. -- 2002. -- Vol. 35, no. 18. -- P. R147--R193.
%
\bibitem{Obzori4} Popov A.~M., Tikhonova O.~V., Volkova
E.~A. Strong-field atomic stabilization: numerical simulation and analytical modelling
// J. Phys.
B.: At. Mol. Opt. Phys. -- 2003. -- Vol. 36, no. 10. -- P. R125--R165.
%
\bibitem{Obzori5} Popov V.~S. Tunnel and multiphoton ionization of atoms and ions in a strong laser field (Keldysh theory)
// Physics-Uspekhi. -- 2004. --
Vol. 47, no. 9. -- P. 855--885.
%
\bibitem{Obzori6} Scrinzi A., Ivanov M.~Yu., Kienberger R., Villeneuve D.~M.
Attosecond physics // J. Phys. B.: At. Mol. Opt. Phys. -- 2006 -- Vol. 39, no. 1. -- P. R1--R37.
%
\bibitem{oscillator1} Efthimiou C.~J., Spector D. Separation of variables and exactly soluble time-dependent potentials in quantum mechanics
// Phys. Rev. A. -- 1994. -- Vol. 49, no. 4. -- P. 2301--2311.
%
\bibitem{oscillator2} Do\^sli\'c N., Danko Bosonac S. Harmonic oscillator with the radiation reaction interaction
// Phys. Rev. A. -- 1995. -- Vol. 51, no. 5. -- P. 3485--3494.
%
\bibitem{Kramers} Henneberger W.~C. Perturbation Method for Atoms in Intense Light Beams
// Phys. Rev. Lett. -- 1968. -- Vol. 21, no. 12. -- P. 838--841.
%
\bibitem{volkov}Wolkow D.~M. \"Uber eine Klasse von L\"osungen der Diracschen Gleichung
// Z. Phys. -- 1935. -- Vol. 94, nos. 3--4. -- P. 250--260.
%
\bibitem{magnus}Magnus W. On the exponential solution of differential equations for a linear operator
// Comm. Pure and Appl. Math. -- 1954. -- Vol. 7. -- P. 649--673.

\end{thebibliography}
\end{document}